\newif\ifsingle

\documentclass[a4paper,conference]{IEEEtran}
\IEEEoverridecommandlockouts

\usepackage[top=1.9cm, bottom=4.3cm,left=1.3cm,right=1.35cm]{geometry}

\usepackage[T1]{fontenc}
\usepackage[latin9]{inputenc}
\usepackage{color}
\usepackage{amsmath}
\usepackage{amssymb}
\usepackage{graphicx}

\makeatletter

\providecommand{\tabularnewline}{\\}

\usepackage[noadjust]{cite}

\ifsingle
\newcommand{\figWidth}{0.65\columnwidth} 
\else
\newcommand{\figWidth}{1\columnwidth} 
\fi 

\definecolor{NewColor}{rgb}{0.2,0,0.5}

\makeatother

\begin{document}
\title{Deep Neural Network-Based Detector for Single-Carrier Index Modulation NOMA}
\author{Toan Gian, Vu-Duc Ngo, Tien-Hoa Nguyen, Trung Tan Nguyen, and Thien Van Luong


\thanks{
Toan Gian, Vu-Duc Ngo, and Tien-Hoa Nguyen are with the School of Electrical and Electronics Engineering, Hanoi University of Science and Technology, Hanoi, Vietnam. (e-mail: toandinh7176@gmail.com,  \{hoa.nguyentien,duc.ngovu \}@hust.edu.vn.} 

\thanks{Trung Tan Nguyen is with the Faculty of Radio-Electronics, Le Quy Don Technical University, Ha Noi 11355, Vietnam (e-mail: trungtannguyen@mta.edu.vn).}

\thanks{Thien Van Luong is with the Faculty of Computer Science, Phenikaa University, Hanoi 12116, Vietnam (e-mail: thien.luongvan@phenikaa-uni.edu.vn).}

\vspace{-0.5cm}
 }

\maketitle
\vspace{-1cm}

\begin{abstract}
In this paper, a deep neural network (DNN)-based detector for an uplink single-carrier index modulation non-orthogonal multiple access (SC-IM-NOMA) system is proposed, where SC-IM-NOMA allows users to use the same set of sub-carriers for transmitting their data  modulated by the sub-carrier index modulation technique. More particularly, users of SC-IM-NOMA simultaneously transmit their SC-IM data at different power levels which are then exploited by their receivers to perform successive interference cancellation (SIC) multi-user detection. The existing detectors designed for SC-IM-NOMA, such as the joint maximum-likelihood (JML) detector and the maximum likelihood SIC-based (ML-SIC) detector, suffer from high computational complexity. To address this issue, we propose a DNN-based detector whose structure relies on the model-based SIC for jointly detecting both $M$-ary symbols and index bits of all users after trained with sufficient simulated data. The simulation results demonstrate that the proposed DNN-based detector attains near-optimal error performance and significantly reduced runtime complexity in comparison with the existing hand-crafted detectors.
\end{abstract}


\vspace{-0.2cm}

\section{Introduction}
 Wireless technologies have been anticipated to move from the conventional orthogonal to non-orthogonal solutions to meet requirements of future wireless networks, such as, ubiquitous coverage, high data rate, low latency, high transmission reliability and scalability ~\cite{Ding2017SurveyNOMA}, \cite{Liu2017SurveyNOMA}. In particular, using non-orthogonal multiple access (NOMA) techniques, different users can simultaneously share the same wireless resources for supporting a massive number of devices. Moreover, in NOMA, multiple users can transmit data at different power levels over the same radio resources, which facilitates successive interference cancellation (SIC) to be effective for detecting the desired signal. Thus, NOMA is a potential solution to improve spectrum efficiency, reduce latency and provide high reliability, and facilitate massive connection compared to the conventional orthogonal multiple access (OMA) \cite{Basar2013ofdmIM}.

Index modulation (IM) has been known as an effective solution to provide trade-off among transmission reliability, spectrum and energy efficiency. In IM-based systems, information bits are implicitly encoded in indices of several facilities such as antennas, subcarriers, and spreading codes \cite{ThienTVT2018}. For example, using the indices of active sub-carriers as an additional domain to the traditional modulated symbols, the orthogonal frequency division multiplexing with IM (OFDM-IM) was proposed in \cite{Basar2013ofdmIM}, which significantly improves the error performance compared with conventional OFDM. Since then, many efforts have been made in order further improve either the error performance or spectral efficiency of OFDM-IM, as can be found in \cite{ThienTVT2018, Mao2017dmIM, ThienTWC2018, wen2017enhancedIM}. Moreover, the bit error rate (BER), symbol error probability (SEP), and outage probability of OFDM-IM using different detection types over different channel state information (CSI) conditions were intensively analyzed in \cite{Luong2018impact,thienWCL2018, thien2017MCIK, Pout2017}, which provide helpful insights into its error performance as well as impacts of CSI uncertainty.

Recently, OFDM-IM has been applied to multi-user communications, particularly to massive machine-type communications (mMTC), which require massive connectivity with highly power-efficient transmissions \cite{Manco2019M2M}. For example, aiming at reducing peak to average power ratio (PAPR) inherited from the OFDM framework, the single-carrier frequency division multiple access (SC-FDMA) was combined with sub-carrier index modulation in a NOMA manner in \cite{Shahab2020IMnoma}. The resultant scheme called SC-IM-NOMA was demonstrated to achieve higher energy efficiency and better BER performance than its single-carrier IM-based OMA counterparts (called SC-IM) proposed in  \cite{Manco2019M2M}, \cite{Chafii2018scfdmaIM}. Note that the SC-IM schemes proposed in \cite{Manco2019M2M} and \cite{Chafii2018scfdmaIM} are based on OMA, in which each user is allocated a dedicated set of subcarriers for employing SC-IM transmissions to achieve better BER performance than conventional SC-FDMA. Yet, SC-IM is not spectrum-efficient compared with SC-IM-NOMA in \cite{Shahab2020IMnoma}, which allows different users to use the same set of sub-carriers for SC-IM transmissions. As a result, SC-IM-NOMA is particularly appropriate for uplink mMTC systems. However, SC-IM-NOMA, the joint maximum-likelihood (JML) detector and the combination of maximum likelihood (ML) and SIC methods (ML-SIC), suffers from high detection complexity compared to both SC-FDMA and SC-IM receivers. Therefore, our paper appears to tackle this fundamental issue.

In this paper, we propose a deep neural network (DNN)-based detector for the uplink SC-IM-NOMA system, called DeepSIC-IM, which can significantly reduce the complexity, while achieving near-optimal BER performance compared with the existing hand-designed detectors. Consider two uplink NOMA users, we design a DNN structure for DeepSIC-IM, that relies on the model-based SIC procedure for jointly detecting both $M$-ary symbols and index bits of both users. More particularly, the proposed DeepSIC-IM consists of two different DNNs, each is dedicated for detecting the signal of the corresponding user when fed with the received signal and the output of the interfering users. Using simulated dataset, DeepSIC-IM is trained offline to jointly minimize the BERs of both users. Then, the trained model can be deployed in an online manner with very low runtime. For this, simulation results are provided to demonstrate that our DeepSIC-IM detector provides near-optimal BER performance at remarkably lower runtime complexity than the existing hand-crafted detectors.

We note that deep learning (DL) or DNN techniques have been widely exploited for enhancing the performance of wireless communication systems. For example, in \cite{Thien2019DLIM}, a DL-based detector was proposed for OFDM-IM, achieving low runtime complexity and near-optimal error performance. The channel estimation and signal detection in OFDM were performed based on DNN in \cite{Ye2018Power}. An application of DL for downlink NOMA receivers was presented in \cite{Luong2022downlink}, which, however, only considers the conventional $M$-ary modulation, without any index modulation. A number of other DL applications for designing the reliable and low-complexity transceiver of wireless systems can be found in \cite{Luong2020engery,Chao2022tubro, Luong2020MC-AE,Luong2022optical}.

The rest of this paper is organized as follows. Section~\ref{sec:System-Model}
describes the system model of SC-IM-NOMA and its existing detection. Section~\ref{sec:proposedDeep}
presents the proposed DeepSIC-IM structure and its corresponding training process. Section~\ref{sec:results} provides simulation results and discussion for BER and runtime complexity. Finally, Section~\ref{sec:consuc} concludes the paper.


\vspace{-0.1cm}

\section{System Model\label{sec:System-Model}}

\vspace{-0.1cm}

\begin{figure}[tb]
\begin{centering}
\includegraphics[width=\figWidth]{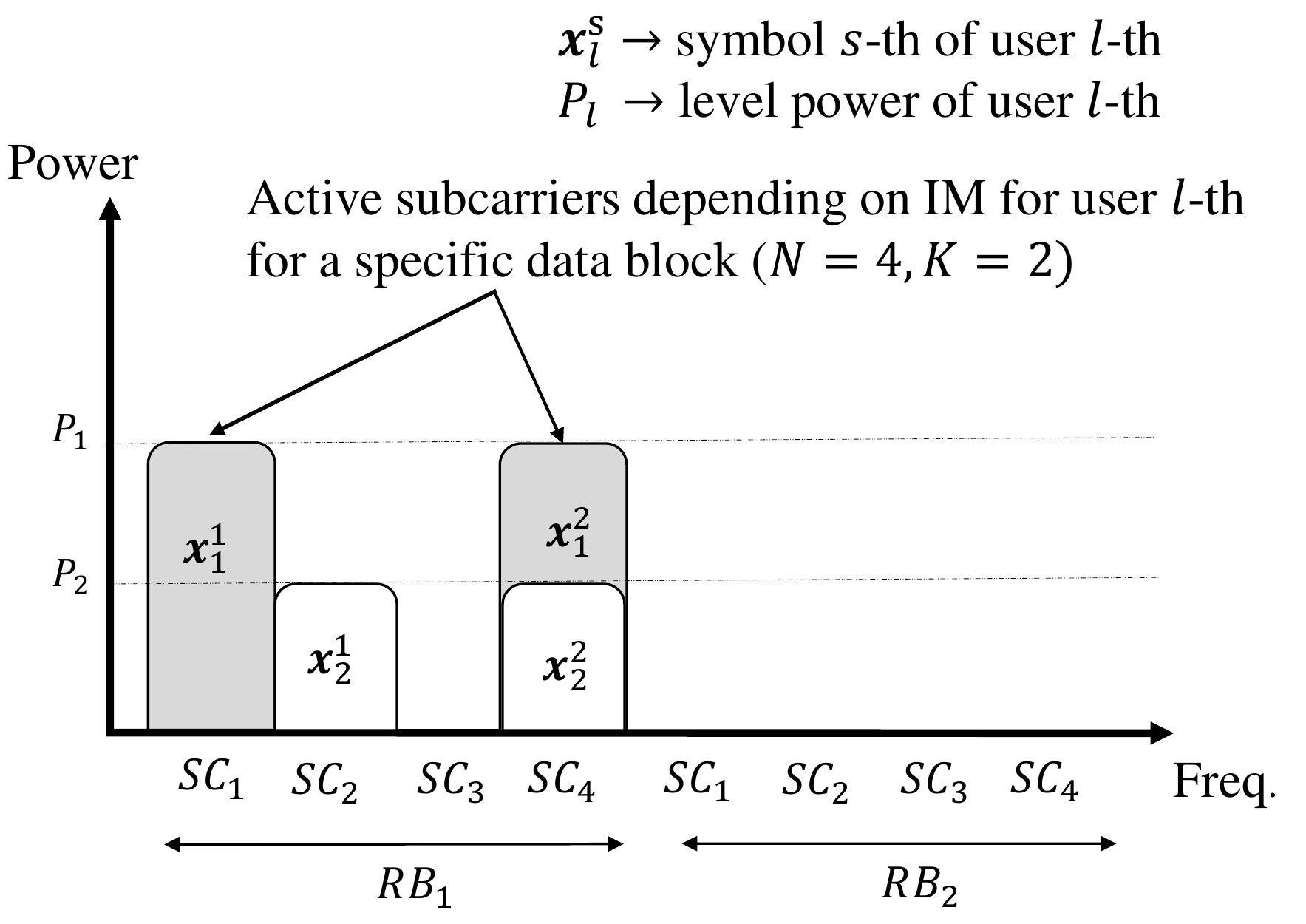} 
\par\end{centering}
\caption{Sub-carrier and power allocation of SC-IM-NOMA.\label{fig:fig1}}
\end{figure}

Consider an uplink scenario where two users $\left(L=2\right)$, $l\in\left(1,2\right)$, transmit data bits to the base station (BS) across the resource block (RB). Gains of channel user 1 to BS and user 2 to BS are represented as $\left|\mathbf{h}_{1}\right|$ and $\left|\mathbf{h}_{2}\right|$, respectively. Both of users transmit SC-IM-based signals to BS with power allocation coefficients of ${P}_{1}$ and ${P}_{2}$, respectively, where $P_{1} > P_{2}$. Each data block takes $b$ bits from each of $L$ users. Additionally, $N$ and $K$ stand for the total number of allotted and active subcarriers, respectively. Each SC-IM block transmits a $b$-bit stream and bits determine the set of $M$-ary modulated symbols that will be conveyed by $K$ active subcarriers. In this paper, all users use the same $N$, $K$ and $M$-ary modulation technique. Thus, $b=b^{ind}+b^{sym}$ = $\left[\log_{2}\left(N,K\right)\right]$ + $K$$\log_{2}$$\left(M\right)$ is the total number of bits transmitted by the $l$-th user in the SC-IM block \cite{Chafii2018scfdmaIM}, \cite{Manco2019M2M}. In the IM subblock, the $l$-th user's initial bits are divided into two parts. The first $b^{ind}=\log_{2}\left(N,K\right)$ bits are to select $K$ active indices using either the combinatorial method or the look-up table \cite{Basar2013ofdmIM}. The remaining $b^{sym}=K\log_{2}\left(M\right)$ bits are mapped to transmitted $M$-ary amplitude/phase modulated (APM) symbols. For example (as Fig.~\ref{fig:fig1}) with $\left(N,K,M\right)=\left(4,2,4\right)$, the transmitted signal vector, $\mathbf{x}_{l}=\left[x_{l}^{1}\,0\,0\,x_{l}^{2}\right]^{T}$, is made up by symbols from an $M$-point constellation and zeros associated with the inactive subcarriers, where $x_{l}^{s}$ stands for the $s$-th symbol of the $l$-th user, $l={1,2}$. The model can serve a certain number of users since SC-IM allocates a set of subcarriers to each user. Fig.~\ref{fig:fig1} depicts a sample SC-IM-NOMA resource mapping for the scenario of two users. Both of users transmit data over the same RB using NOMA at distinct power levels. 

The received signal at the BS over $N$ subcarriers can be represented by
\begin{equation}
\mathbf{y}=\sum_{l=1}^{L}\sqrt{P_{l}}\mathbf{h}_{l}\odot\mathbf{x}_{l}+\mathbf{w},\label{eq:Rate}
\end{equation}
where $\odot$ stands for the element-wise multiplication, $\mathbf{y}=\left[y\left(1\right)\,,\ldots\,,y\left(N\right)\right]^{T}$, $\mathbf{h}_{l}=\left[{h}_{l}\left(1\right),\ldots,{h}_{l}\left(N\right)\right]^{T}$ is the channel vector, $\mathbf{x}_{l}=\left[x_{l}\left(1\right)\,,\ldots\,,x_{l}\left(N\right)\right]^{T}$ and $\mathbf{w}=\left[w_{1}\,, \ldots\,, w_{N}\right]^{T}$ is the additive white Gaussian noise (AWGN) vector, $P_{l}$ is the power for the $l$-th user.

In order to recover transmitted signal, the JML detector and ML-SIC detector \cite{Shahab2020IMnoma} can be employed, which are described in the following.

\subsubsection{JML detector} The JML detector performs an exhaustive search to simultaneously recover the transmitted vector $\mathbf{x}_{l}$  for all $l$-th users. Consider the scenario of two users, both transmitted vectors $\mathbf{x}_{1}$ and $\mathbf{x}_{2}$ are recovered simultaneously as follow:
\begin{equation}
\left\{ \mathbf{\hat{x}}_{l}\right\} _{l=1,2}=\underset{\mathbf{x}_{l=1,2}\in\delta}{\text{arg min}}\left\Vert \mathbf{y}-\sqrt{P_{1}}\mathbf{h}_{1}\odot\mathbf{x}_{1}-\sqrt{P_{2}}\mathbf{h}_{2}\odot\mathbf{x}_{2}\right\Vert, \label{eq:JML_equation}
\end{equation}
where JML checks $\mathbf{y}$ for all possible combinations of transmit vectors of users, which is denoted by $\mathbf{\delta}$. This detector can provide the optimal performance at the expense of high computational complexity, thus, it can be used as a benchmark when designing any low complexity receiver.

\subsubsection{ML-SIC detector} Based on this, the SIC-ML detector of user $l$-th operates in the following iterative fashion. In the case of 2 users, user 1, which is allocated more power, will detect its own signal first by treating the interference from the user 2 as the noise. Using the ML criterion, the estimated signal of user 1 is determined by
\begin{equation}
\hat{\mathbf{x}}_{1}=\underset{\mathbf{x}_{1}\in\mathbf{\delta}}{\text{arg min}}\left\Vert \mathbf{y}-\sqrt{P_{1}}\mathbf{h}_{1}\odot\mathbf{x}_{1}\right\Vert .\label{eq:SIC-equation}
\end{equation}

Then, the contribution of user 1 to $\mathbf{y}$ is eliminated for decoding the signal of user 2. Explicitly, the symbol of user 2 is recovered using the ML estimate in which the interfering signal of user 1 is estimated by $\mathbf{\hat{x}}_{1}$, resulting in
\begin{equation}
\hat{\mathbf{x}}_{2}=\underset{\mathbf{x}_{2}\in\mathbf{\delta}}{\text{arg min}}\left\Vert \left(\mathbf{y}-\sqrt{P_{1}}\mathbf{h}_{1}\odot\hat{\mathbf{x}}_{1}\right)-\sqrt{P_{2}}\mathbf{h}_{2}\odot\mathbf{x}_{2}\right\Vert .\label{eq:SIC2-equation}
\end{equation}

From the formula \eqref{eq:JML_equation}, \eqref{eq:SIC-equation} and \eqref{eq:SIC2-equation}, the complexity of the conventional detectors mentioned above would increase, particularly when having more uplink NOMA users as well as when the SC-IM-NOMA parameters increase. Motivated by this issue, we propose a DL-based detector to detect transmitted signals with competitive performance and low complexity compared to these ML-based detectors.

\vspace{-0.1cm}

\section{ Proposed DeepSIC-IM Detector\label{sec:proposedDeep}}

\vspace{-0.1cm}
In this section, the structure of the proposed DeepSIC-IM detector and its training procedure are presented.

\vspace{-0.2cm}

\subsection{Structure of proposed  DeepSIC-IM\label{subsec:3a}}

\vspace{-0.1cm}
\begin{figure}[tb]
\begin{centering}
\includegraphics[width=\figWidth]{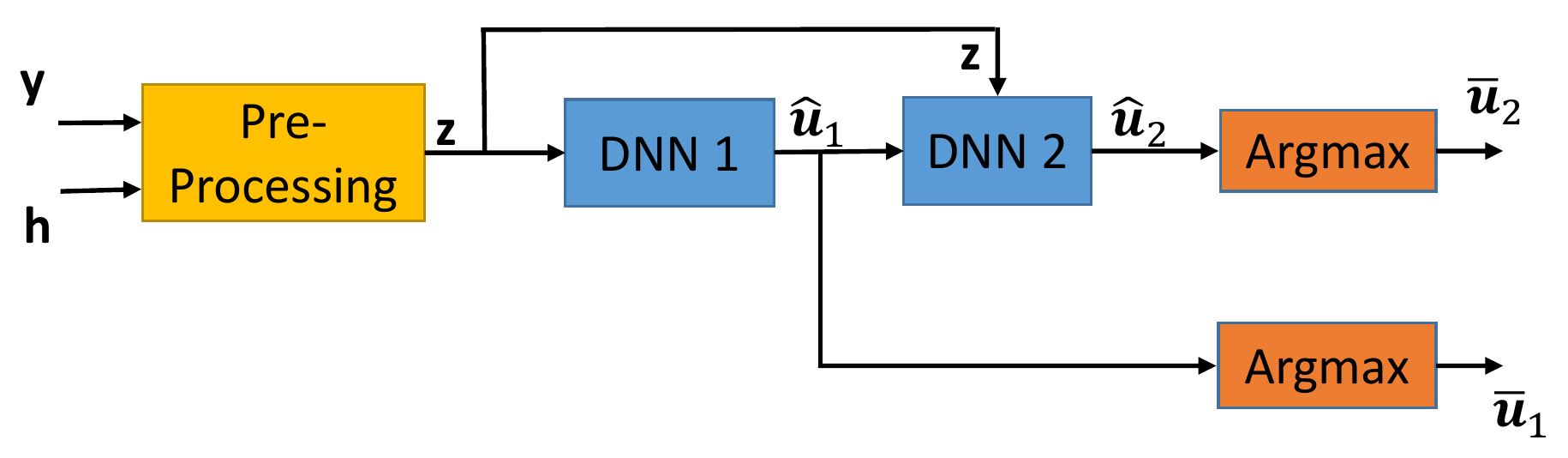}
\par\end{centering}
\caption{The proposed DeepSIC-IM DNN structure.\label{fig:fig2}}
\end{figure}

The structure of DeepSIC-IM is illustrated in Fig.~\ref{fig:fig2}. Assume that the channel state information (CSI) is perfectly known at the receiver and considered as input of the decoder, along with the received signal $\mathbf{y}$. Hence, the well-known zero-forcing (ZF) equalizer is employed to get an equalized received signal vector as $\mathbf{\bar{y}}=\mathbf{y}\odot \mathbf{h}^{-1}_{1}$. Particularly, the complex vectors $\mathbf{\bar{y}}_{R}$ and $\mathbf{\bar{y}}_{I}$ are transformed into an $2N$-dimensional real vector $\mathbf{z}=\left[\mathbf{\bar{y}}_{R},\mathbf{\bar{y}}_{I}\right]$, as shown in Fig.~\ref{fig:fig2}, where $\mathbf{\bar{y}}_{R}$, $\mathbf{\bar{y}}_{I}$ are real and imaginary parts of $\mathbf{\bar{y}}$.
\begin{table}
\caption{\textcolor{black}{Parameters of DNN 1 and DNN 2 with $\left(N,K,M\right)=\left(4,1,4\right)$}\label{tab:table1}}
\centering{}
\begin{tabular}{|l|c|c|}
\hline 
Parameter & DNN 1 & DNN 2 \tabularnewline
\hline 
\hline 
Input size & $2N+\left(l-1\right)2^{b}$= 8 & $2N+\left(l-1\right)2^{b}$= 24\tabularnewline
\hline 
Output size & $2^{b}$= 16 & $2^{b}$= 16\tabularnewline
\hline 
FC layer & 3 & 3\tabularnewline
\hline 
Hidden Nodes & 32-64 & 128-256\tabularnewline
\hline 
\end{tabular}
\end{table}

The input of the $l$-th DNN block includes both $\mathbf{z}$ and the output of the previous $l-1$ DNN blocks, namely $\mathbf{\text{s}}_{l}$. More specifically, those elements are concatenated to form an input vector of the size $\left[2N+\left(l-1\right)2^{b}\right]$. Due to the complexity of each signal being different (the signal detection for $l$-th user will be more difficult than that for $\left(l-1\right)$-th user), each DNN to detect each signal will have a different number of nodes, as shown in Table~\ref{tab:table1}. 

The proposed decoder of each user has three nonlinear FC layers. At the hidden layer, either the rectifier linear unit (Relu), $f_{\text{Relu}}\left(x\right)=\text{max}\left(0,x\right)$, or the hyperbolic tangent (Tanh) function, $f_{\text{Tanh}}(x)\,=\,\frac{1-e^{-2x}}{1+e^{-2x}}$, can be used as the activation function. The decoder generates an output probability vector $\mathbf{\hat{u}}_{l}$ with the Softmax layer as the output layer. The output vector $\mathbf{\hat{u}}_{l}$ of each DNN with size $\left[2^{b}\right]$ can be given~by
\begin{equation}
\mathbf{\hat{u}}_{l}=f_{\text{Softmax}}\left(\mathbf{W}_{3}f_{\text{Relu}}\left(\mathbf{W}_{2}f_{\text{Relu}}\left(\mathbf{W}_{1}\mathbf{\mathbf{s}}_{l}+\mathbf{b}_{1}\right)+\mathbf{b}_{2}\right)+\mathbf{b}_{3}\right),\label{eq:outputDNN}
\end{equation}
where $\mathbf{\text{W}}_{1}$, $\mathbf{\text{b}}_{1}$, $\mathbf{\text{W}}_{2}$, $\mathbf{\text{b}}_{2}$ and $\mathbf{\text{W}}_{3}$, $\mathbf{\text{b}}_{3}$ denote the weights and biases of the first, second and third FC layers, respectively, $\mathbf{s}_{l}$ is the input vector of the $l$-th DNN, which is represented by
$\mathbf{s}_{l}=\left[\mathbf{z},\mathbf{\hat{u}}_{1},\ldots,\mathbf{\hat{u}}_{l-1}\right]$. Finally, we get the vector $\bar{\mathbf{u}}_{l}$ based on the largest entry of $\hat{\mathbf{u}}_{l}$. Then, the information prediction will be calculated.

Crucial points of the proposed scheme structure can be described as follows. First, the length of input and output of DeepSIC-IM is determined by system parameters such as $N$, $K$ and $M$. This will make the complexity of the proposed scheme highly dependent on each of the three aforementioned indicators. Intuitively, when the number of transmitted $b$-bit increases, we need a number of nodes large enough to guarantee a predetermined performance. Therefore, the number of nodes of the $l$-th DNN (denoted by $\mathcal{O}_{l}$) must be carefully chosen to obtain the desired performance for each system configuration. The value of $\mathcal{O}_{l}$ must be sufficiently large to ensure a predetermined performance. In addition, we can attain a promising trade-off between the detection accuracy and the model complexity by adjusting value of $\mathcal{O}_{l}$.

\vspace{-0.2cm}

\subsection{Training Procedure \label{subsec:3b}}
\vspace{-0.1cm}
 Before implementing the proposed DeepSIC-IM technique, the DNN model with the collected data from simulation need to be trained. Initially, various $b$-bit sequences will be randomly created for $L$ users and transformed into the matching vector $\mathbf{x}_{l}$. The one-hot vectors $\mathbf{u}_{l}$ are generated first to create $L$ sets of one-hot labels for $L$ users, while noise vectors are randomly generated. At the receiver, $\mathbf{y}$ and $\mathbf{h}_{1}$ are pre-processed to obtain the vector $\mathbf{z}$, which is then combined with the output of the previous $l-1$ DNN blocks to provide the input for the $l$-th DNN block $\mathbf{s}_{l}$. The mean square error (MSE) cost function which is adopted to measure the difference between the estimation vector $\mathbf{\hat{u}}_{l}$ and its true value vector $\mathbf{u}_{l}$, as follows:
 \begin{equation}
\mathcal{L}\left(\mathbf{u}_{l},\hat{\mathbf{u}_{l}};\mathbf{\theta}_{l}\right)=\sum_{l=1}^{L}\frac{1}{b}\left\Vert \mathbf{u}_{l}-\hat{\mathbf{u}}_{l}\right\Vert ^{2},\label{eq:lostfunc}
\end{equation}
where $\theta_{l}=\left\{ \mathbf{W}_{i},\mathbf{b}_{i}\right\} _{i=1,2,3}$ are the weights and biases of the $l$-th DNN user model, $\mathbf{u}_{l}$ is the collection of one hot vector corresponding to $b$ bits trained offline and $\mathbf{\hat{u}}_{l}$ is the output vector of the $l$-th DNN. Model parameters are updated for the batches randomly picked up from data samples, using the stochastic gradient descent (SGD) algorithm as follows
 \begin{equation}
\theta^{+}:=\theta-\eta\nabla\mathcal{L}\left(\mathbf{u}_{l},\mathbf{\hat{u}}_{l};\theta\right),\label{eq:SGD}
\end{equation}
where $\eta$ is learning rate.  In our training, we use the adaptive moment estimation (Adam) optimizer, a sophisticated updating technique built on SGD. The signal to noise ratio (SNR) training (denoted as $\lambda_{train}$), which has a significant impact on the performance of the training model, will be chosen to suit the model as closely as possible. For instance, if $\lambda_{train}$ train is too small, the training will not account for the influence of noise, which will result in poor generalization of the trained model. Choosing a suitable $\lambda_{train}$ for achieving the desired performance will be presented in detail in the simulations.

\section{Simulation results\label{sec:results}}

\vspace{-0.1cm}

This section presents simulation results to demonstrate the performance of our DeepSIC-IM compared to the existing schemes, including JML and ML-SIC detectors, in terms of both BER and runtime complexity.

\begin{table}
\caption{A summary of simulation parameters\label{tab:para}}

\centering{}%
\begin{tabular}{|l|r|}
\hline 
Parameter  & Value\tabularnewline
\hline 
\hline 
Number of users $L$  & 2\tabularnewline
\hline 
Parameters of index modulation ($N$, $K$, $M$)  & 4,1,4\tabularnewline
\hline 
Gains of channel $\mathbf{h}_{1}$, $\mathbf{h}_{2}$ & $\left[2,2,2,2\right]$, $\left[1,1,1,1\right]$\tabularnewline
\hline 
Power allocation coefficient $P_{1}$,$P_{2}$  & 2,1\tabularnewline
\hline 
Hidden nodes of DNN 1  & 32-64\tabularnewline
\hline 
Hidden nodes of DNN 2  & 128-256\tabularnewline
\hline 
Activation for hidden layers of DNN 1  & \textcolor{black}{Tanh}\tabularnewline
\hline 
Activation for output layers of DNN 1, 2 & \textcolor{black}{Softmax}\tabularnewline
\hline
Activation for hidden layers of DNN 2  & \textcolor{black}{Relu}\tabularnewline
\hline 
Training SNR $\lambda_{train}$  & 16, 18, 20, 22, 24 dB\tabularnewline
\hline 
Learning rate $\eta$  & 0.001\tabularnewline
\hline 
Batch size  & 200\tabularnewline
\hline 
Number of training epochs  & 500\tabularnewline
\hline 
Training data size  & $2\times10^6$\tabularnewline
\hline 
Testing data size  & $10^{6}$\tabularnewline
\hline 
Optimizer  & Adam \cite{Kingma2014AdamAM} \tabularnewline
\hline 
\end{tabular}
\end{table}

\subsection{Parameter Setting}
\label{subsec:para-setting}

In Table~\ref{tab:para}, we summarize the SC-IM-NOMA system parameters as well as the DNN model parameters, which are used for conducting simulations in this section. For example, we train our DeepSIC-IM with 500 epochs, each contains $10^{4}$ samples, where the batch size is 200 samples. During training, data samples are randomly generated for each batch. Therefore, we have a total of $2\times10^{6}$ different data samples for training, while the testing size is $10^{6}$ samples. The learning rate $\eta$ is set to be 0.001. 

We note that in this work we only consider fixed channels for both users, as shown in Table~\ref{tab:para}. For varying or random fading channels, we may have to use some advanced training techniques, such as feed-forward error correction (FEC)-aided online labeling \cite{Luong2022downlink}. We consider such the extension as our future work. Furthermore, throughout extensive experiments, we discovered that user 1 would prefer the Tanh activation function, while user 2 would prefer the Relu activation function, for better BER performance, as shown in Table~\ref{tab:para}.

\subsection{BER Performance}
\label{subsec:berperformance}

\begin{figure}[tb]
\begin{centering}
\includegraphics[width=\figWidth]{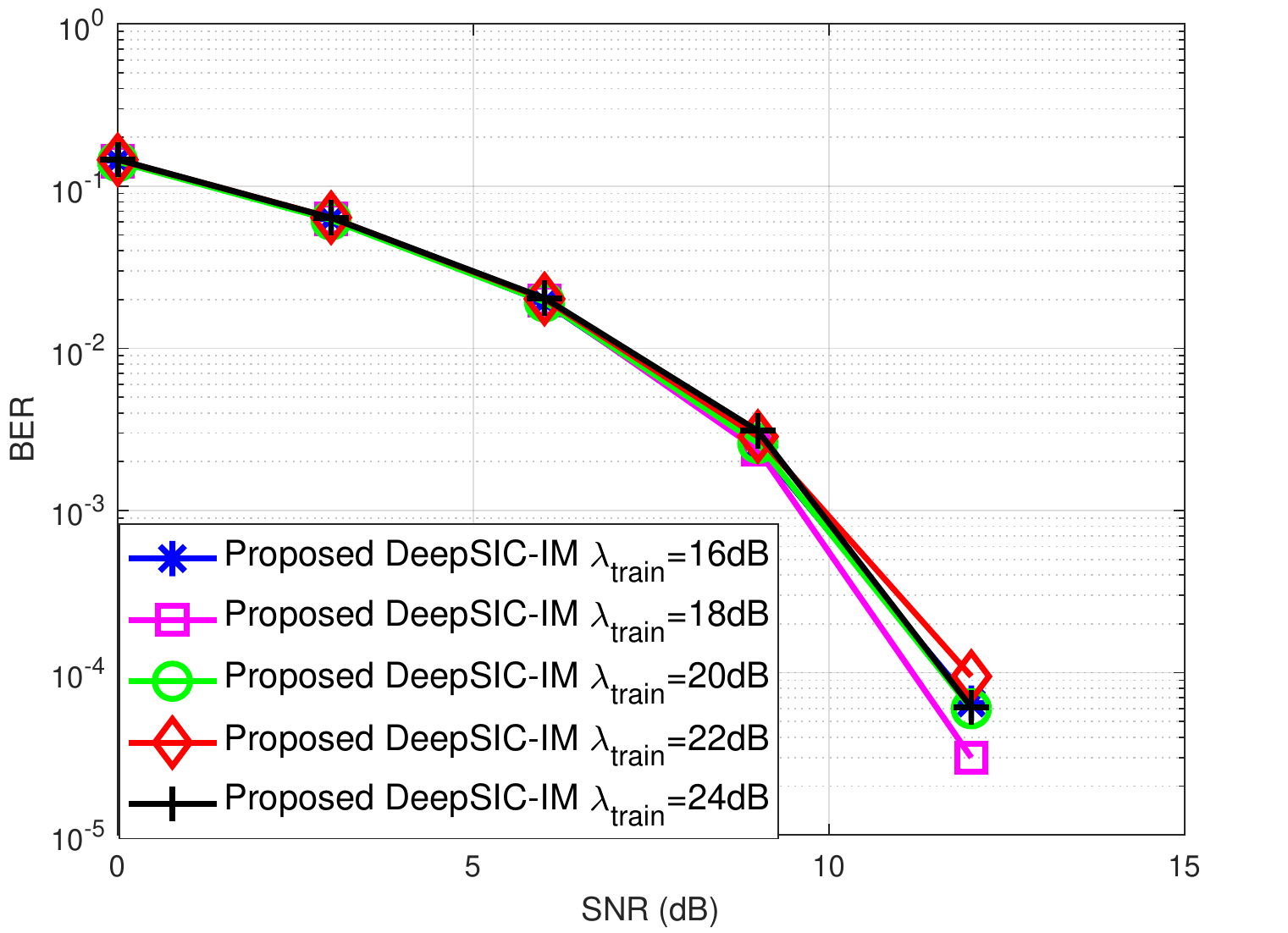}
\par\end{centering}
\caption{BER performance of DeepSIC-IM for user 1 trained at various training SNRs $\lambda_{train}$. The training setting is described in Table~\ref{tab:para}.\label{fig:ber1}}
\end{figure}

\begin{figure}[tb]
\begin{centering}
\includegraphics[width=\figWidth]{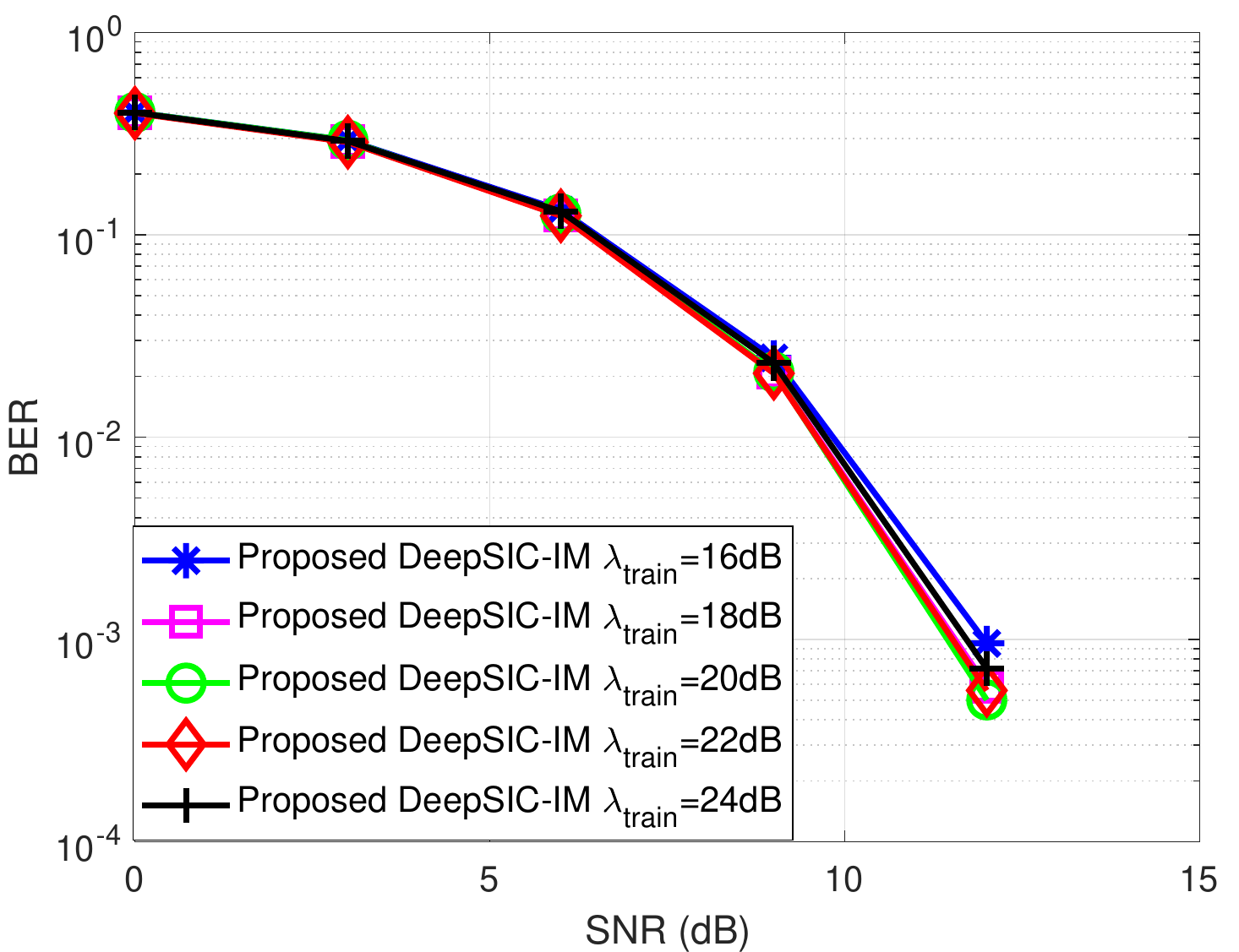}
\par\end{centering}
\caption{BER performance of DeepSIC-IM for user 2 trained at various training SNRs $\lambda_{train}$. The training setting is described in Table~\ref{tab:para}.\label{fig:ber2}}
\end{figure}

The BER performance of user 1 and user 2 using the proposed DeepSIC-IM trained at various training SNRs is illustrated in Fig.~\ref{fig:ber1} and Fig.~\ref{fig:ber2}, respectively. It is shown from both figures that the BER performance achieved by DeepSIC-IM for both users varies with respect to the training SNR, especially at high testing SNRs. It is also observed that our proposed DeepSIC-IM tends to achieve the best BER for the two users when the training SNR is between 18 and 20~dB. Therefore, in the following, we will use a training SNR of 18 dB for training DeepSIC-IM to compare with the baseline.

\begin{figure}[tb]
\begin{centering}
\includegraphics[width=\figWidth]{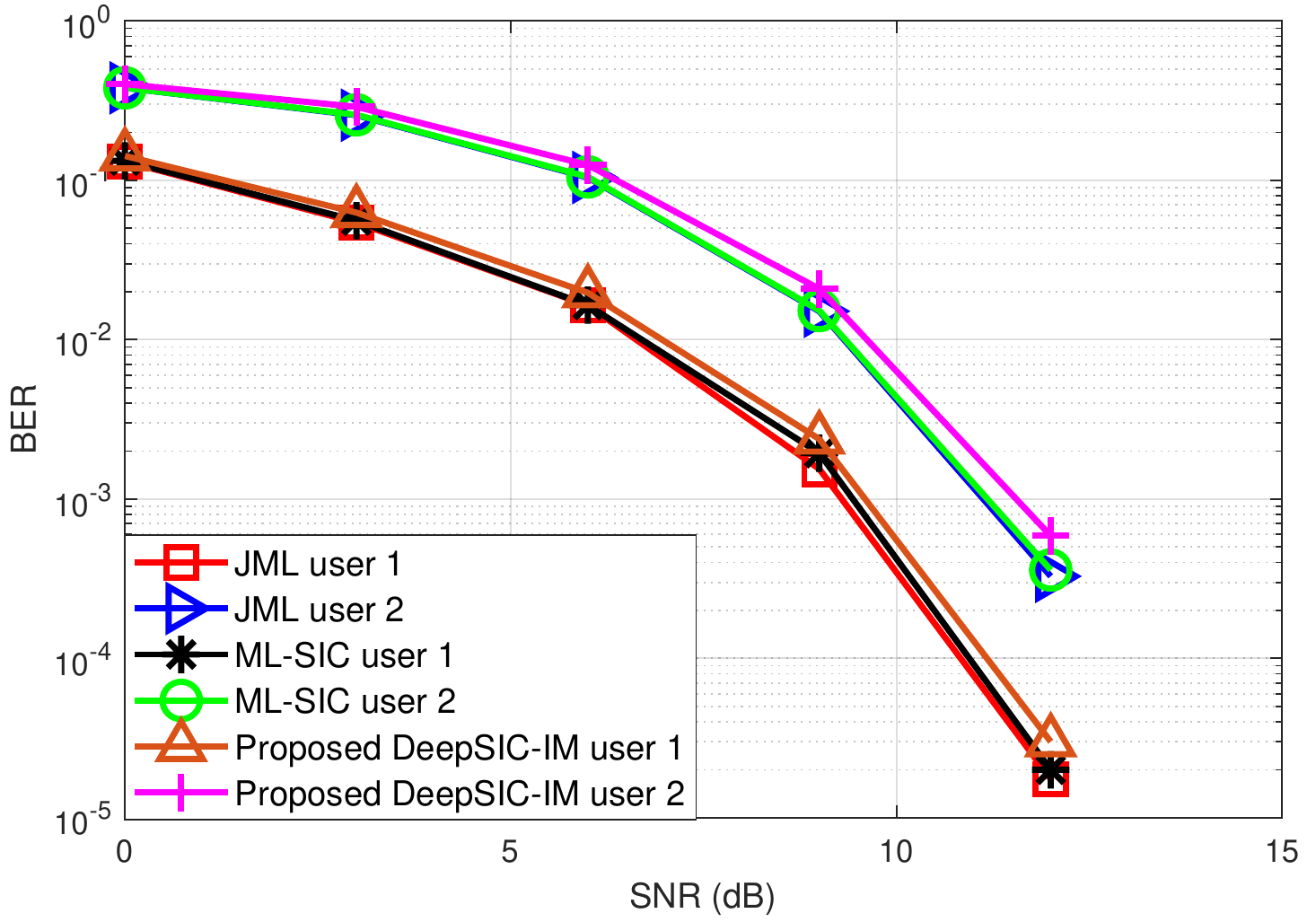}
\par\end{centering}
\caption{BER comparison between the proposed DeepSIC-IM and baselines with training SNR $\lambda_{train}$ = 18dB. The training setting is described in Table~\ref{tab:para}.\label{fig:comber}}
\end{figure}


Fig.~\ref{fig:comber} compares the BER performance among the proposed DeepSIC-IM detector, JML and ML-SIC detector, where both users are considered, and our scheme is trained at 18~dB. It is shown from Fig.~\ref{fig:comber} that the proposed learning detector attains BER performance very close to that of the JML and ML-SIC detectors. For example, at the BER of $10^{-2}$, the SNR gap between the proposed detector, JML and ML-SIC is negligible with less than 0.4 dB and 0.3 dB, respectively, while our detector requires much lower runtime complexity than its counterpart, as shown in the following subsection.

\subsection{Runtime Complexity}
\label{subsec:runtime}

We now measure the runtime for detecting each sample of the proposed DeepSIC-IM, JML and ML-SIC detectors. In particular, we convert the DeepSIC-IM model that was trained using the Tensorflow library to MATLAB. To ensure a fair comparison, the JML and ML-SIC detectors are also run on MATLAB on the same computer. The obtained complexity for the three detectors measured in seconds is compared in Table~\ref{tab:comrun}. It is show from this table that the proposed learning detector demands much less runtime than both the JML and ML-SIC detectors. More particularly, the runtime of our DeepSIC-IM is 7 and 3 times less than that of the JML and ML-SIC, respectively, which clearly validates the advantage of the proposed detector in terms of runtime complexity.

\begin{table}[!ht]
\centering
\caption{Complexity comparison among DeepSIC-IM, JML and ML-SIC\label{tab:comrun}}

\begin{tabular}{|l|c|c|r|}
\hline 
$(N,K,M)$ & JML & ML-SIC & DeepSIC-IM\tabularnewline
\hline 
\hline 
(4,1,4) &$7.65\times10^{-4}$  &$3.31\times10^{-4}$  &$1.14\times10^{-4}$ \tabularnewline
\hline 
\end{tabular}

\end{table}

\vspace{-0.2cm}
 
\section{Conclusions\label{sec:consuc}}

\vspace{-0.1cm}
In this paper, we proposed the DeepSIC-IM detector for the SC-IM-NOMA system. In particular, we designed a novel DNN structure for SC-IM-NOMA that includes several DNN blocks, each used for detecting the transmitted signal of the corresponding user, where its input data is made up of the pre-processed received signal and the output of the interfering users. Once trained with simulated dataset, our proposed method can be deployed for detecting data bits in an online manner with very low runtime. Our simulation results demonstrated that the proposed DeepSIC-IM detector achieves near-optimal BER performance at remarkably lower runtime complexity compared to the existing ML-based detectors. Although we only consider two uplink NOMA users in this work, the extension of DeepIM-SIC for more users is straightforward. In the future, we plan to extend the proposed DeepSIC-IM to time-varying channels, where the channel coding-aided online training technique in \cite{Luong2022downlink} is helpful.

\vspace{-0.1cm}

 \bibliographystyle{IEEEtran}
\bibliography{IEEEabrv,refs}

\end{document}